\documentclass[aps,twocolumn,showpacs,superscriptaddress,floatfix]{revtex4-1}
\usepackage{graphicx,amsfonts,amssymb,amsmath,mathrsfs,hyperref,bm,physics}

\usepackage{epstopdf}

\newif\ifhyper
\hypertrue
\ifhyper
\hypersetup{
   citecolor = {red},
   colorlinks = {true}, 
   linkcolor = {blue},
   urlcolor = {blue} 
}
\fi

 \makeatletter
    \renewcommand\@make@capt@title[2]{%
     \@ifx@empty\float@link{\@firstofone}{\expandafter\href\expandafter{\float@link}}%
      {\textbf{#1}}\@caption@fignum@sep#2\quad}%
    \makeatother
   
\makeatletter 
\renewcommand{\fnum@figure}{\textbf{Figure~\thefigure}}
\makeatother

\begin{document}

\title{Numerical and quantum simulation of a quantum disentangled liquid}

\author{E. Abbasgholinejad}
\affiliation{Department of Physics, Sharif University of Technology, P.O.Box 11155-9161, Tehran, Iran}
\affiliation{Department of Electrical and Computer Engineering, University of Washington, Seattle, WA 98195, USA }

\author{S. Raeisi}
\affiliation{Department of Physics, Sharif University of Technology, P.O.Box 11155-9161, Tehran, Iran}
\email{email}

\author{A. Langari}
\affiliation{Department of Physics, Sharif University of Technology, P.O.Box 11155-9161, Tehran, Iran}
\email{langari@sharif.edu}

\begin{abstract}
The illustrative wave function for a quantum disentangled liquid (QDL) composed of light and heavy particles
is examined within numerical simulations.  
Initial measurement on light particles gives rise to the volume law of the entanglement entropy of the heavy particles subsystem.
The entropy reaches its maximum value as the ratio of the system to subsystem sizes increases. The standard deviation
of entanglement entropy from its thermodynamic limit due to the initial configuration of the light particle is diminished 
within ensemble averaging. We have introduced a quantum circuit to simulate the underlying
QDL state. The results of the quantum simulation are in agreement with the numerical simulations which confirms that the introduced circuit realizes a QDL state.
\end{abstract}

\pacs{75.10.Jm}

\maketitle
\section{Introduction}
\label{introduction}
Thermalization or the approach to thermal equilibrium in isolated many-body quantum systems is a fundamental problem in quantum statistical physics \cite{Deutsch:1991,Srednicki:1994,Rigol:2008}. There are various quantum systems, whose properties can be described
on the basis of equilibrium statistical mechanics, while there are also systems in which the thermal equilibrium is not reached.
For instance, integrable models or systems which exhibit many-body localization 
\cite{Anderson:1958,Kramer:1993,Gornyi:2005,Basko:2006,Oganesyan:2007,Pal:2010,Bauer:2013,Imbrie:2016}
due to strong disorder, represent
non-thermalized phase \cite{Luca:2016,Abanin:2017,Alet:2018}.
Recent theoretical studies have revealed that integrability and the existence of a static
disorder are not necessary conditions \cite{Muller:2015,Yao:2016,Smith:2017,Yarloo:2018,Yarloo:2019}
for the violation of thermalization, 
while there are translationally invariant
and non-integrable systems that do not thermalize in the conventional sense, among them is the Quantum Disentangled Liquid (QDL) \cite{Grover}.

QDL is a finite temperature phase of translationally invariant quantum liquid consisting of two 
(or more) species of indistinguishable particles with a
large mass ratio, namely heavy and light ones. The basic characteristic of the QDL phase is that, while the heavy degrees
of freedom are fully thermalized, the light ones satisfy an area law of entanglement entropy for a typical fixed
configuration of the heavy particles. This suggests that the light particles are ``localized'' by their heavy 
counterparts.
Thus, in a QDL phase, thermal equilibration is incomplete and it is not a fully ergodic phase, where we are faced with the partial breakdown of thermalization in a translationally invariant system. 

The playground is a closed quantum system, which evolves in time and its dynamics is given by Hamiltonian 
via the Schr\"{o}dinger equation. The time evolution operator is unitary, which guarantees no loss of information
for the whole state of the system. However, the underlying information is distributed to all parts of the system according
to the time evolution process, which changes the entanglement between different parts of the system. In this respect,
part of a closed quantum system is immersed in a heat bath of the rest of the system, which opens the opportunity to localize a subsystem while the rest is in a thermalized phase.
A qualitative diagnostic to identify the QDL phase
is the bipartite entanglement entropy after a projective measurement of the heavy/light degrees of freedom. 
The entanglement entropy scales with the volume of the system (volume law) in a thermalized phase, while it grows
at most proportional to the area of the system (area law) for a non-thermalized phase.
Proposals for QDL state have been introduced in bosonic model \cite{Muller:2015},  spin-charge degrees
of freedom in a Hubbard-like model \cite{Garrison:2017} and also anyons of the Kitaev ladder \cite{Yarloo:2018}.

In this work, we examine the QDL state introduced in Ref.\cite{Grover}. We present a quantum simulation scheme for QDL state \cite{cirac2012goals, georgescu2014quantum}. 
Specifically, we introduce a quantum circuit for implementing the QDL state on a quantum simulator. 
First, we use numerical simulation to justify the 
QDL properties of the introduced state and investigate its characteristics on a finite-size system.
We observe that the measurement of light particles leaves the heavy particles in a thermalized phase, where the 
entanglement entropy obeys the volume law. However, in a finite-size system entanglement entropy is attained for
a large ratio of the size of the heat bath to its subsystem. An ensemble average justifies that our results do 
not depend on the initial configuration of light particles. The inverse process, measurement of heavy particles, 
leads to zero entanglement entropy of light particles, which verifies the area law for the light species.
We compare the results of numerical simulations with their counterparts on a quantum simulator. Although the number of qubits in the quantum simulator is limited, our results justify that the introduced quantum circuit truly 
produce a QDL state.
A QDL state can be considered as a pragmatic system for quantum bits, which should be interesting to realize 
a quantum memory \cite{Grover,Yao2016}.

In the next section, we review the notion of QDL and establish the state introduced in Ref.\cite{Grover}.
Then, our proposed quantum circuit is introduced. The numerical and quantum simulations are presented in 
Sec.\ref{qs}. Finally, we summarize and conclude our results.

\section{Quantum Disentangled Liquid}
\label{qdl}

\subsection{Illustrative wavefunction}
\begin{figure}
  \centering
  \begin{tabular}{@{}c@{}}
    \includegraphics[width=0.45\columnwidth]{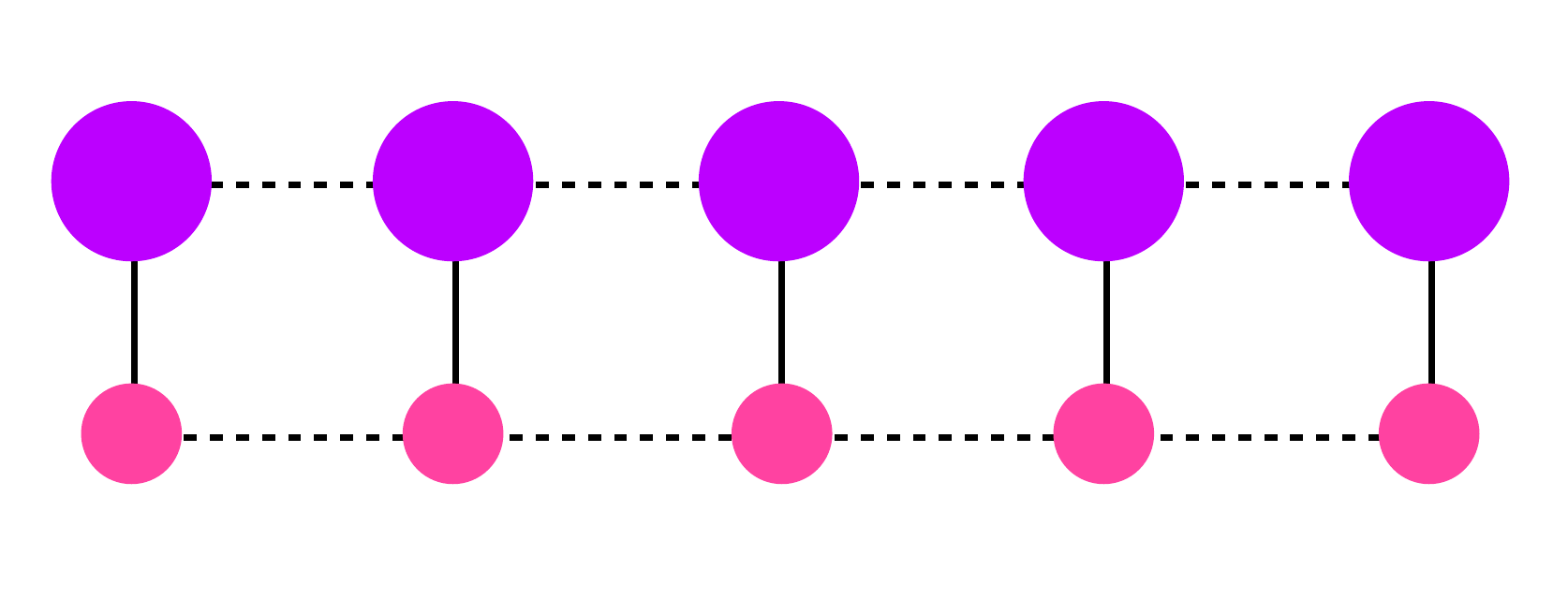} \\[\abovecaptionskip]
    \small (a) 
  \end{tabular}
  \vspace{\floatsep}
  \begin{tabular}{@{}c@{}}
    \includegraphics[width=0.45\columnwidth]{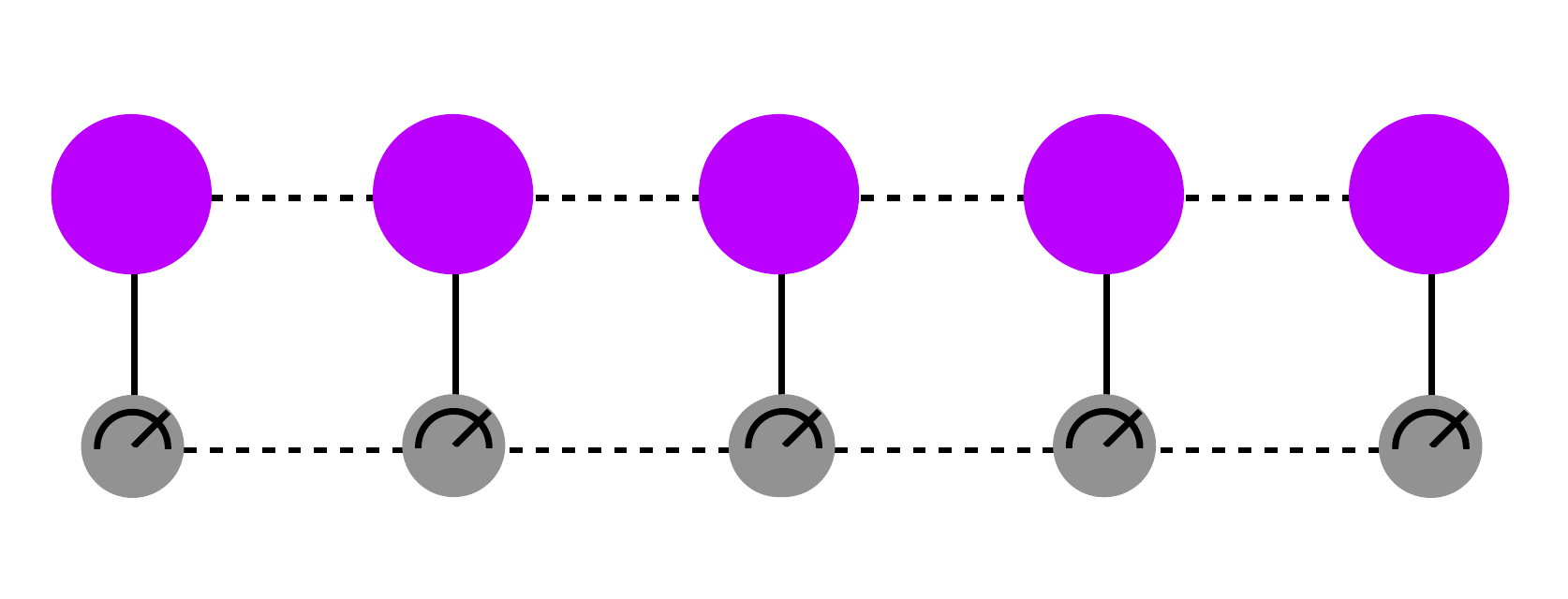} \\[\abovecaptionskip]
    \small (b) 
  \end{tabular}
  \vspace{\floatsep}
  \begin{tabular}{@{}c@{}}
    \includegraphics[width=0.45\columnwidth]{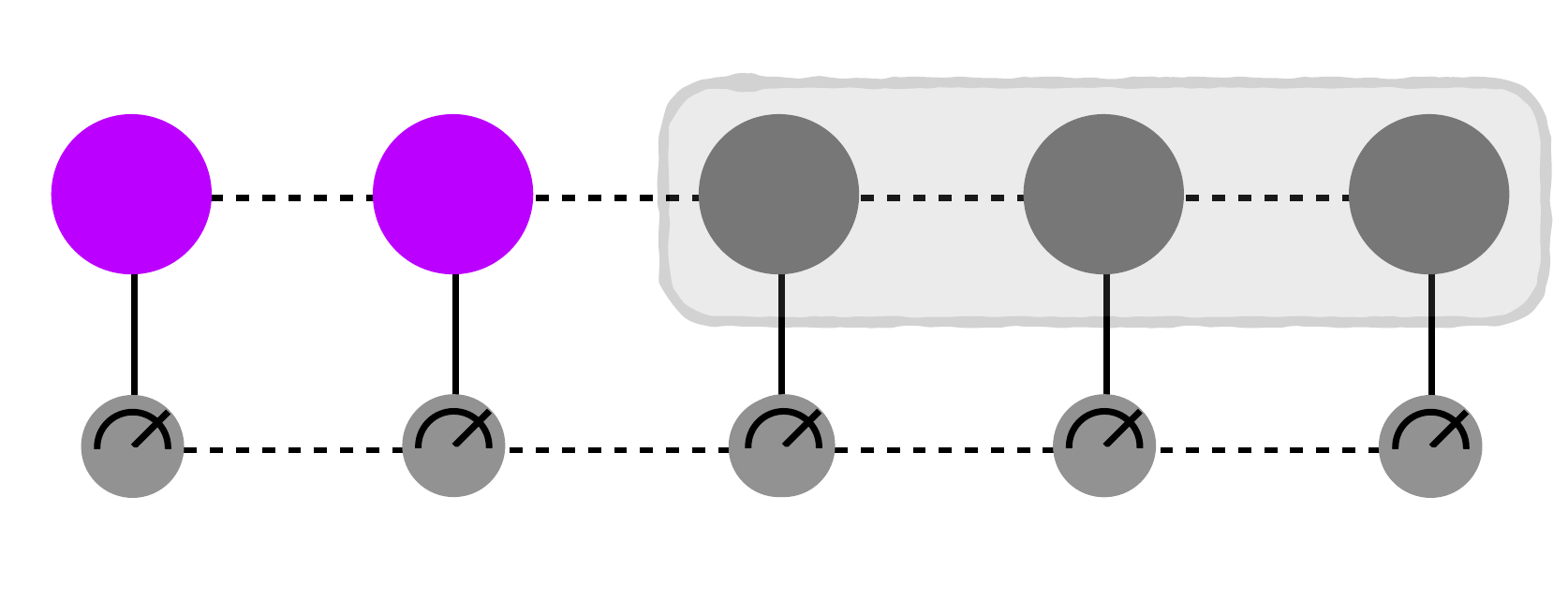} \\[\abovecaptionskip]
    \small (c) 
  \end{tabular}  
   \begin{tabular}{@{}c@{}}
    \includegraphics[width=0.45\columnwidth]{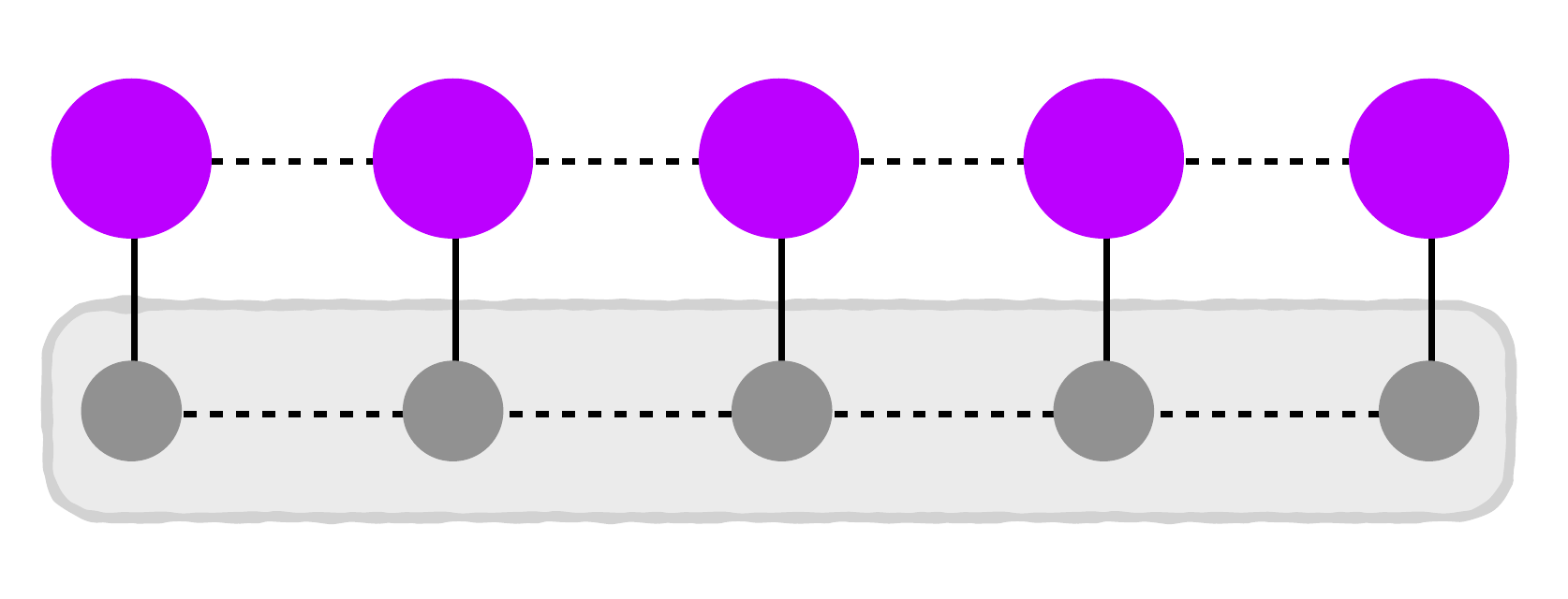} \\[\abovecaptionskip]
    \small (d) 
  \end{tabular}  
  \caption{(a) Schematic representation of Light (small circles) and heavy (large circles) particles in the system. 
 (b) Measurement is done on light particles.
 (c) Tracing out a subsystem of heavy particles after measurement on light particles. 
 (d) Tracing out light particles in the initial system.}
  \label{visual1}
\end{figure}

Grover and Fisher introduced an illustrative wave function for a lattice model consisting of two types of heavy and light particles, which represents the state of quantum disentangled liquid \cite{Grover}. 
The proposed state can be understood in terms of the notion of {\it many-body localization}. In an extreme limit, where the position of heavy particles are fixed without any dynamics, the pattern of heavy particle positions can be considered as a random background for the light ones. This randomness is a disordered space for the light particles and could lead to the localization of light particles. On the other hand, if we only think about heavy particles a random configuration of heavy particles would be a result of thermalized state. 
The system is composed of ${\cal N}$ unit cells, each containing a pair of heavy and light particles
as shown in Fig.\ref{visual1}-(a).
In the framework of occupation number representation, the wavefunction is written as:
\begin{equation}
    \ket{\psi }_{QDL} = \sum_{\{N_j\}, \{n_j\}} \psi(\{N_j\}, \{n_j\}) \, \ket{\{N_j\}}\ket{\{n_j\}},
	\label{gro1}
\end{equation}
where $\{ N_j=0, 1; j=1,\dots, {\cal N}\}$ , $\{ n_j=0, 1; j=1, \dots, {\cal N}\}$ are the occupation number of heavy and light particles, respectively.
The coefficients are given by the following expression
\begin{equation}
     \psi(\{N_j\}, \{n_j\})  = \psi(\{N_j\}) \prod_{j=1}^{{\cal N}} \frac{1}{\sqrt{2}}[\delta_{n_j,0} + e^{i\pi N_j}\delta_{n_j,1} ].
     \label{gro2}
\end{equation}
The summation in Eq. (\ref{gro1}) is over all possible $4^{\cal N}$ configurations. The first term on the right hand side of 
Eq.(\ref{gro2}), namely random phase, is a function of heavy particles alone, 
\begin{equation}
\psi(\{N_j\}) = \frac{1}{\sqrt{2^{{\cal N}}}} sgn (\{N_j\}), 
\label{randonsign}
\end{equation}
where $sgn(\{N_j\}) = \pm 1$ is a random sign, chosen independently for all of $2^{\cal N}$ heavy particles configurations. Moreover, the second part of Eq. (\ref{gro2}), namely occupation phase, could be categorized into four different configurations as explained below,
\begin{equation}
\begin{cases}
    N_j=\bold{0},n_j=0 \, : \, \delta_{n_j,0} + e^{i\pi N_j}\delta_{n_j,1}=1 ,\\ 
    N_j=\bold{0},n_j=1 \, : \, \delta_{n_j,0} + e^{i\pi N_j}\delta_{n_j,1}=1 ,\\ 
    N_j=\bold{1},n_j=0 \, : \, \delta_{n_j,0} + e^{i\pi N_j}\delta_{n_j,1}=1 ,\\ 
    N_j=\bold{1},n_j=1 \, : \, \delta_{n_j,0} + e^{i\pi N_j}\delta_{n_j,1}=-1 .\\ 
\end{cases}
\label{st}
\end{equation}
As a result there is only a minus sign due to occupation of both heavy and light particles, simultaneously. 

Consider a system consisting of $2{\cal N}$ qubits, the first ${\cal N}$ qubits represent the heavy particles and the rest represents light particles at different sites. 
As an example, for ${\cal N}=2$ the Hilbert space bases are given below
\begin{equation}
\begin{aligned}
    \{ \ket{\{N_j\}} \} &= \{ \ket{\bold{00}}, \ket{\bold{01}},\ket{\bold{10}},\ket{\bold{11}} \}, \\
    \{ \ket{\{n_j\}} \} &= \{ \ket{00}, \ket{01},\ket{10},\ket{11} \} .
\end{aligned}
\end{equation}
Hence,  using the configurations given in Eq. (\ref{st}), the QDL wavefunction is written in the following form
\begin{equation}
\begin{aligned}
    \ket{\psi }_{QDL} = \frac{1}{2^2} &\Big[ \\
      sgn (\{\bold{00}\}) \big( &\ket{\bold{00},00} + \ket{\bold{00},01} + \ket{\bold{00},10} +\ket{\bold{00},11} \big) \\ 
    +sgn (\{\bold{01}\}) \big( &\ket{\bold{01},00} -\ket{\bold{01},01} +\ket{\bold{01},10} -\ket{\bold{01},11} \big)  \\ 
    +sgn (\{\bold{10}\}) \big( &\ket{\bold{10},00} +\ket{\bold{10},01} -\ket{\bold{10},10} -\ket{\bold{10},11}  \big) \\  
    +sgn (\{\bold{11}\}) \big( &\ket{\bold{11},00} -\ket{\bold{11},01} -\ket{\bold{11},10} +\ket{\bold{11},11}  \big)  
    \Big] .
\end{aligned}
\label{wf}
\end{equation}
We emphasize that the first two qubits are heavy particles at each site and the rest shows light particles. For example $\ket{\bold{01},01}$ means that the first site has no particles and the second one is occupied with both a heavy and a light particle, which leads to a minus sign as a consequence of the second site.


\subsection{Quantum circuit}\label{mqc}

\begin{figure*}
\includegraphics[width=1.9\columnwidth]{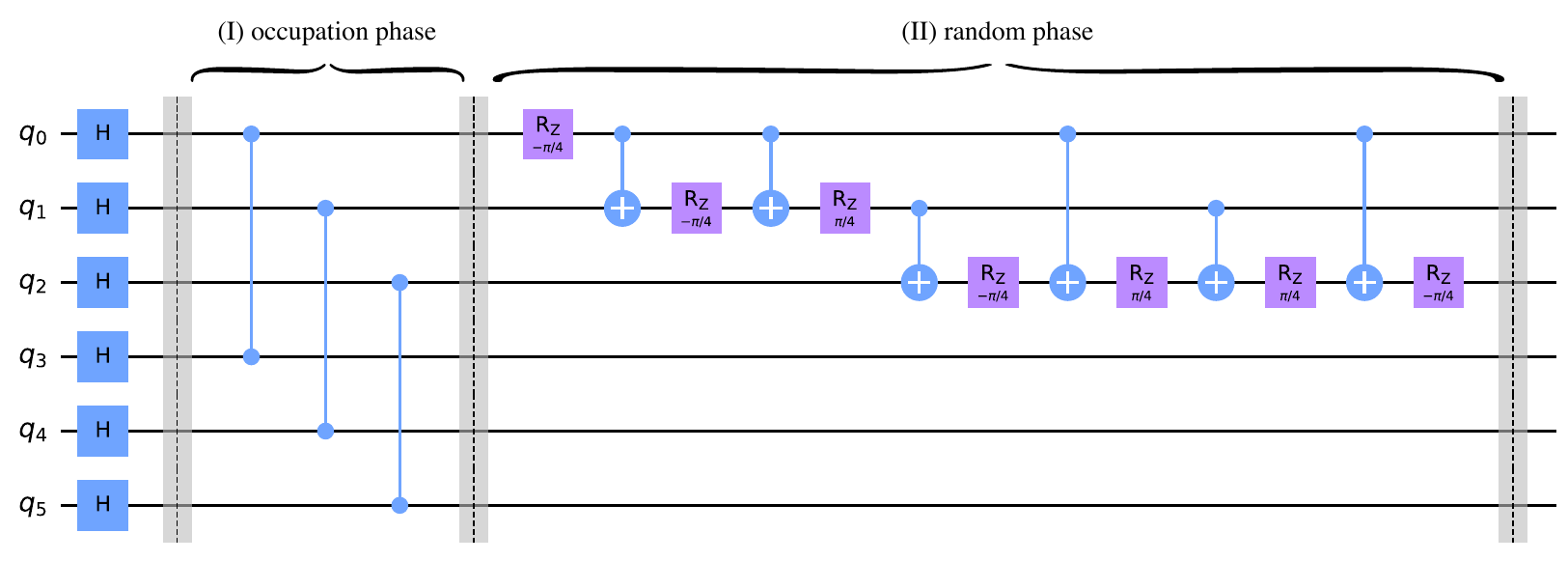}
\caption{The six-qubit quantum circuit for preparing the QDL state with ${\cal N}=3$. The first three qubits,  named ($q_0$, $q_1$, $q_2$), represent the occupation of heavy particles in each site and the qubits  ($q_3$, $q_4$, $q_5$) are used for light particles. All the qubits are initialized in $\ket{0}$ state. Then, the Hadamard gates make a uniform distribution of all the possible basis states. The next step is to set the phases in Eq. (\ref{gro2}). The phase is -1 when both the heavy and light sites are occupied simultaneously. The qubits that are connected by controlled-Z gates share the same site on the lattice. So when both sites are occupied, i.e. $\mid \bold{1}1\rangle$, the states pick up a negative sign, otherwise, it stays unchanged. 
The last step is to apply the random signs. The sign is chosen independently for each of the $2^{\cal N}$ heavy particles configuration. This part of the circuit assigns a random sign to the states of the heavy sites. For more details on this part of the circuit, see Appendix \ref{rpc}.
 }
\label{circ}
\end{figure*}

In this section we introduce an algorithm for preparing the wavefunction explained in the previous section (Eq. (\ref{gro1})). 
First, we prepare a uniform superposition of all the elements of the computational basis. Each element of the superposition is affected by two sets of phases. The first one is the phase that is set by the occupation of light and heavy particles on each site. The second one is the random phase in Eq. \ref{randonsign}. 

In our algorithm, we start with all qubits initialized in $\ket{0}$. Then, we apply Hadamard gates to all of them to prepare the uniform superposition of all elements of the computational basis. This gives

\begin{equation}
      H^{\otimes 2\cal N} \ket{0}^{\otimes 2\cal N} = \frac{1}{2^{\cal N}} \prod_1^{2\cal N}\big( \ket{0} + \ket{1} \big) = \frac{1}{2^{\cal N}} \sum_{i=0}^{2^{2\cal N}-1} \ket{i}. 
\end{equation}
Next we need to add the occupation phases. Specifically, for a given site, when there is both a heavy and a light particle, the state picks up a negative sign. 
This can be implemented using Controlled-Z gates between qubits corresponding to the light and heavy particle of each site. This indicates that we need $\cal N$ Controlled-Z gates for this stage of the algorithm. 


For the last part, we need to add the random phase to each element of the superposition. 
Specifically, we select a random array of $\pm 1$ of length $N$ and apply these random signs to different configurations of heavy particles. Note that the phases only depend on the heavy particles and as a result, this part should only act on the first $\cal N$ qubits. 
This can be done by a diagonal unitary gate (in the computational basis) acting only on the qubits corresponding to the heavy particles. 
The diagonal of the unitary operator would be the random array of $\pm 1$  signs. 
\begin{equation}
      U_{\vec{\theta}} = diag \{ e^{-i\theta_0}, \cdots ,  e^{-i\theta_{m-1}}\}
\end{equation}
To break this operator into single and two-qubit gates, we use the method introduced by Schuch and Siewert in \cite{Schuch}, which is described fully in Appendix \ref{rpc}.  It utilizes the one-qubit gate, phase shift (z-rotation), and the CNOT gate to build any arbitrary $m=2^{\cal N}$ dimensional controlled phase shift. Note that this prepares any random phase and is not limited to $\pm 1$.  The optimal circuit, at worst, is made up of $2^{\cal N}-1$ one-qubit z-rotation gates and $2^{\cal N}-2$ CNOT. Therefore, the depth of the circuit for this part grows exponentially with $\cal N$. 

The quantum circuit for the full algorithm, for three sites, ${\cal N}=3$, is shown in Fig.\ref{circ}. The occupation phases are implemented with the Controlled-Z gates in part I and the random phases are implemented in part II.

\section{Numerical simulation of QDL state}\label{qs}

Here, we examine the properties of a QDL state using numerical simulation at finite system size, ${\cal N}$. 
The QDL state given in Eq.(\ref{gro1}) is realized numerically, assuming a random distribution of signs defined in Eq.(\ref{randonsign}).
Next, we will average over an ensemble. The details of the averaging will be further discussed.

\begin{figure}
\includegraphics[width=0.9\columnwidth]{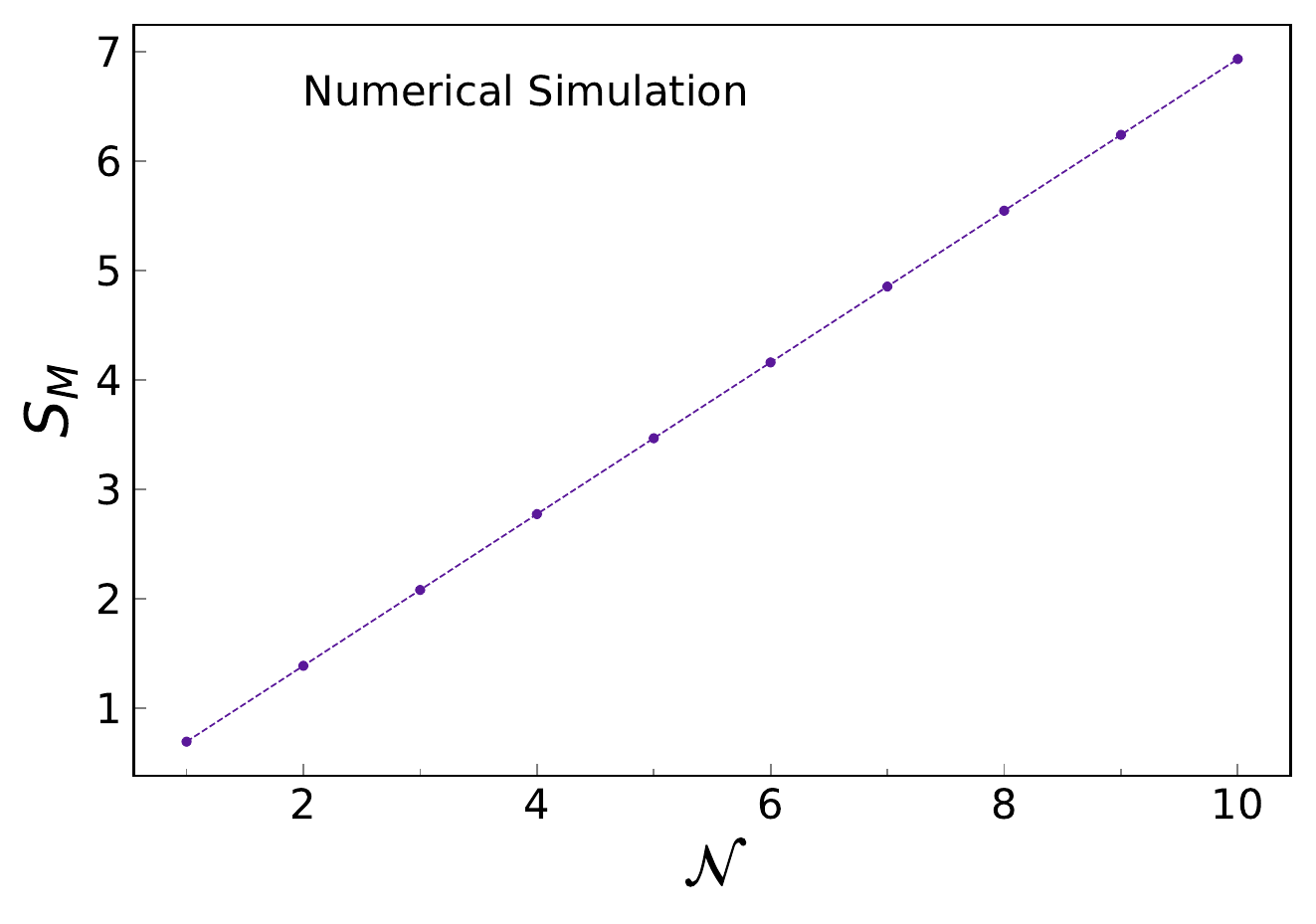}
\caption{R{\'e}nyi-2 entropy of all heavy particles is plotted versus the number of sites. It shows that the inter-component entanglement scales with the volume of the whole system.
}
\label{tot_en}
\end{figure}

Firstly, we would like to see the scaling of entanglement under the partitioning between the two components of particles. In this respect, the light particles' ($m$) degrees of freedom are integrated out from the density matrix ($\rho_Q$), which is 
constructed from the pure state of Eq.(\ref{gro1}). The process is shown schematically in 
Fig.\ref{visual1}-(d).
The resulting reduced density matrix
($\rho_M$)
is given by
\begin{eqnarray}
    \rho_{Q}&=&\ket{\psi} _{QDL\;\; QDL}\bra{\psi}, \\
    \rho_M &=& \trace _m (\rho_{Q}).
\end{eqnarray}
The R{\'e}nyi entanglement entropies can be used to quantify the amount of entanglement
\begin{equation}
S(\alpha)=\frac{1}{1-\alpha} \ln \trace_M[(\rho_M)^{\alpha}],
\label{R-entropy}
\end{equation}
where it reduces to the von Nuemann entropy in the limit of $\alpha \rightarrow 1$.
Here, we use the second R{\'e}nyi  entropy as a measure of entanglement
\begin{equation}
    S_M = - \ln [\trace_M (\rho_M^2)],
\end{equation} 
and call it R{\'e}nyi-2 for simplicity.
The von Neumann entropy is greater than or equal to the R{\'e}nyi-2 entropy, hence, the 
volume law scaling of R{\'e}nyi-2 entropy results in the volume law of the von Neumann entropy.
The numerical simulation of the R{\'e}nyi-2  entropy of the reduced density matrix
$S_M$, is plotted in Fig.\ref{tot_en}, for several values of system sizes. It shows a linear dependence on the number of sites for the entanglement (see the dotted line in Fig.\ref{tot_en}).
The R{\'e}nyi-2  entropy of heavy particles scales as the volume law, which indicates that the
underlying state belongs to an ergodic phase (for heavy particles),

\begin{equation}
    S_{M} = - \ln [\trace_M (\rho_M^2) ] = {\cal N} \ \ln 2  \propto L^d.
\end{equation}

Now, we intend to see the effect of measurement on the QDL state. We split the procedure into two
subsections, measuring light and heavy particles.

\begin{figure}
\includegraphics[width=0.9\columnwidth]{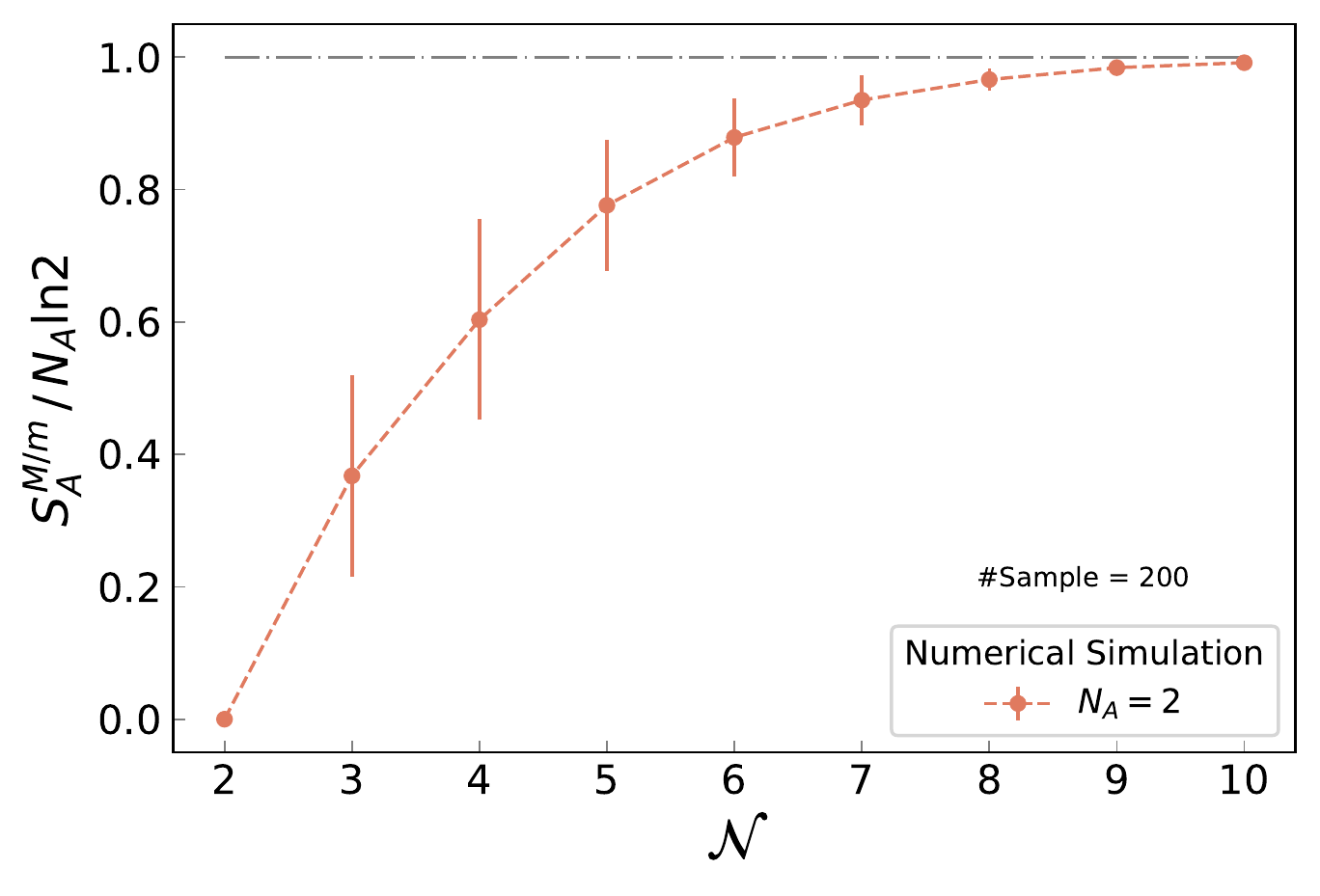}
\caption{R{\'e}nyi-2 entropy of heavy particles after measurement of light particles, versus system size (${\cal N}$) for A-subsystem with a fixed number of sites $N_A=2$. At high ${\cal N}$,  the heavy particles' entropy reaches its maximum,$S^{M/m}_A = 2 \ln 2$, manifesting the volume law, which is a key feature of QDL.  
Ensemble average is taken into account for 200 initial configurations of light particles. }
\label{sim_en}
\end{figure}

\subsection{Measurement of light particles}\label{mlp}

We start with the state of system given by the density matrix, $\rho_Q$. 
By measuring the position of light particles, which has been shown schematically in 
Fig.\ref{visual1}-(b),
the reduced density matrix of heavy particles is obtained, 
\begin{equation}
    \rho(\{ \tilde{n}_j \}) = \bra{\tilde{n}} \rho_Q \ket{\tilde{n}},
    \label{rho_n}
\end{equation}
which is a function of the measured values of $\{ \tilde{n}_j \}$.
Then, we consider a bi-partitioning of system into two regions A(subsystem) and B(environment) with number of sites $N_A$ and $N_B$, respectively, where $N_A + N_B ={\cal N}$. The reduced density matrix of region $A$ is obtained by tracing out heavy particles of region $B$:
\begin{equation}
    \rho_A(\{ \tilde{n}_j \}) = \trace_B \rho(\{ \tilde{n}_j \}),
\end{equation}
which has been depicted in Fig.\ref{visual1}-(c).
The R{\'e}nyi-2  entropy of subsystem A is given by
\begin{equation}
 S_A^{M/m} =  - \ln [\trace_{A} \big(\rho_A(\{ \tilde{n}_j \})^2 \big) ].
 \label{S_AM}
\end{equation}
It should be emphasized that the resulting entropy ($S_A^{M/m}$) is still a function of the
measured position of light particles, which has been randomly chosen for the initial QDL state. To get an impression of the ensemble average of the QDL state, the obtained 
entropy ($S_A^{M/m}$) is averaged over the random locations of light particles, which are assumed to come from a homogeneous distribution. Our result, which has been 
averaged over 200 samples of random configurations of light particles, is plotted
in Fig.\ref{sim_en}. In this figure, the normalized entropy of heavy particles 
is plotted versus the system size (${\cal N}$), where the size of the A-subsystem is 
fixed to $N_A=2$. As Fig.\ref{sim_en} shows, the entropy reaches its maximum for 
large system size (thermodynamic limit), i.e. 
\begin{equation}
    S_A^{M/m} = N_A \ \ln 2  \sim L_A^d.
    \label{S_AM_volume_law}
\end{equation}
This is consistent with the theoretical expectations in Ref.\onlinecite{Grover}.
The above equation (Eq.(\ref{S_AM_volume_law})) represents a volume law scaling, which justifies that measuring light particles does not affect the thermal behavior of heavy ones.
This is a necessary condition of a QDL state.

\begin{figure}
  \centering
    \includegraphics[width=0.95\linewidth]{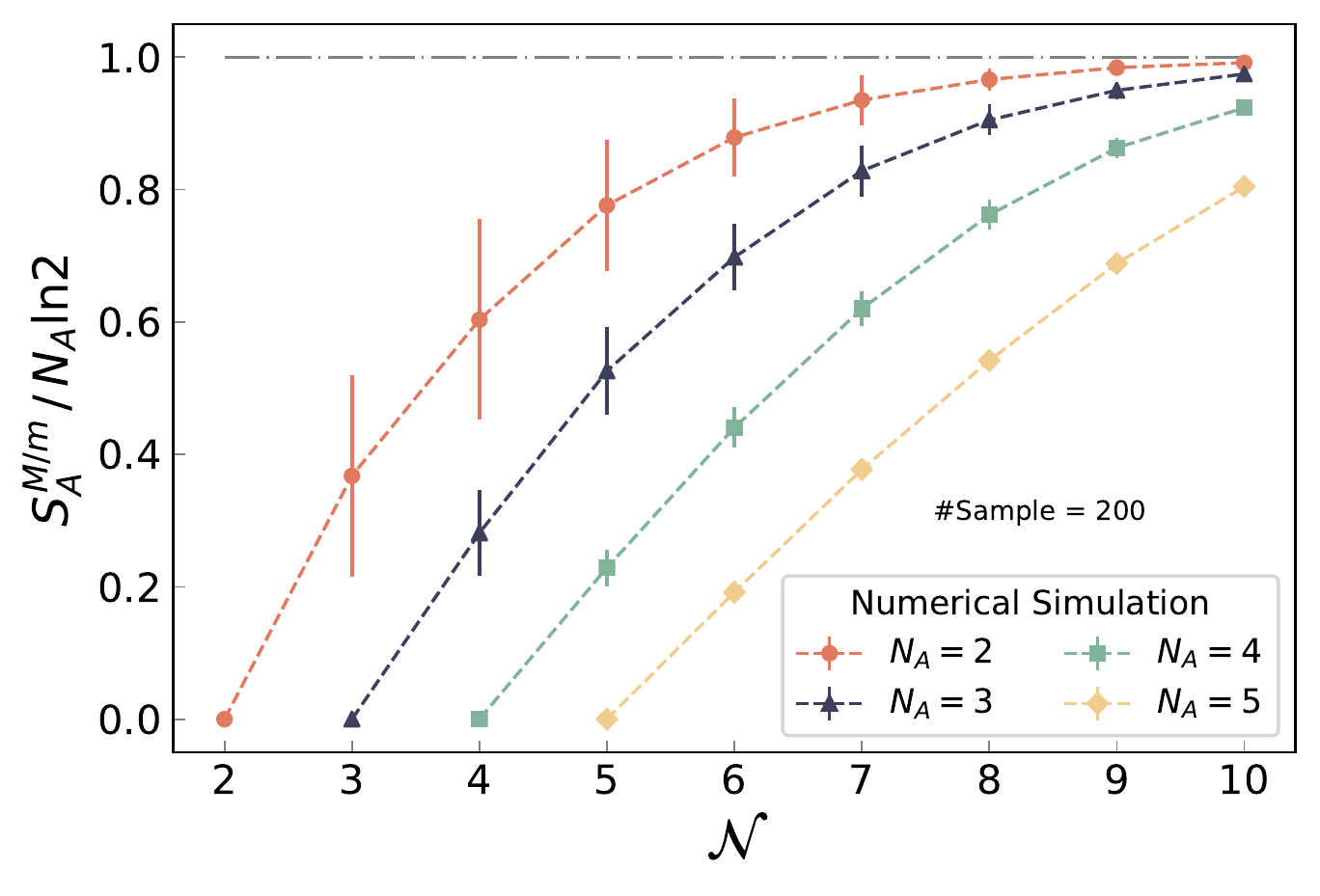}
    \caption{Normalized R{\'e}nyi-2 entropy of heavy particles after measurement of light particles versus system size for several A-subsystem, $N_A=2, 3, 4, 5$. The average on the random initial configurations of light particles is done on 200 samples.
}
    \label{N2345}
    \end{figure}

We have also examined how the size of the A-subsystem affects the entropy of heavy particles.
Fig.\ref{N2345} shows the normalized entropy of heavy particles versus the whole system size, ${\cal N}$, where
different plots show fixed values of A-subsystem size, $N_A=2, 3, 4, 5$. Again, the ensemble average is taken 
over 200 configurations of light particles. All plots confirm that upon increasing system size (${\cal N}$), the entropy reaches its maximum value. However, for a larger A-subsystem, the maximum entropy is attained for larger size of the whole system, while the error bar of ensemble average is smaller for larger $N_A$. 
The qualitative behavior of all plots is the same and shows similar trends.

\begin{figure}
  \centering
  \begin{tabular}{@{}l}
       \small (a) \\
    \includegraphics[width=0.95\columnwidth]{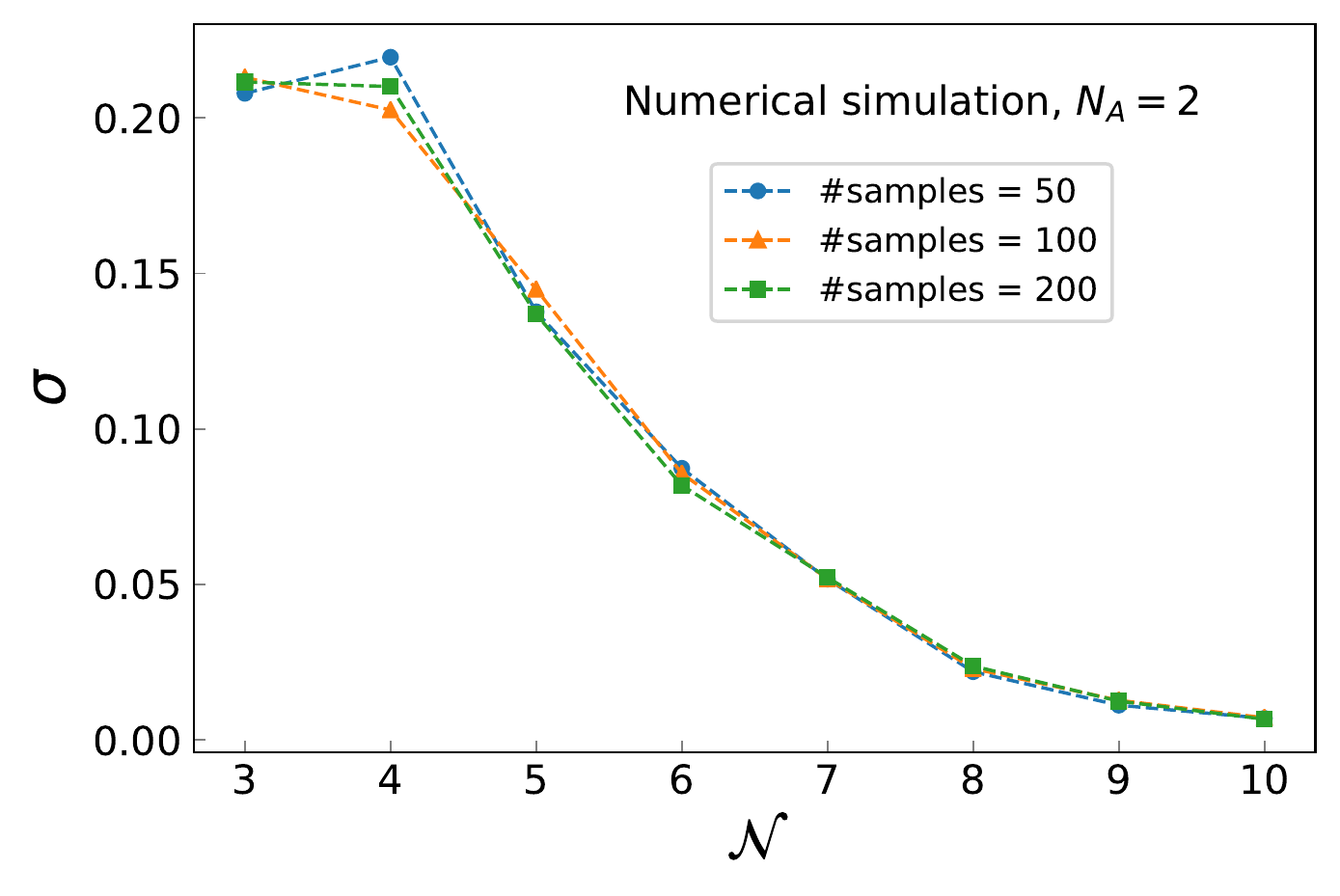} \\
  \end{tabular}\\
  \vspace{\floatsep}
  \begin{tabular}{@{}l}
            \small (b) \\
    \includegraphics[width=0.95\columnwidth]{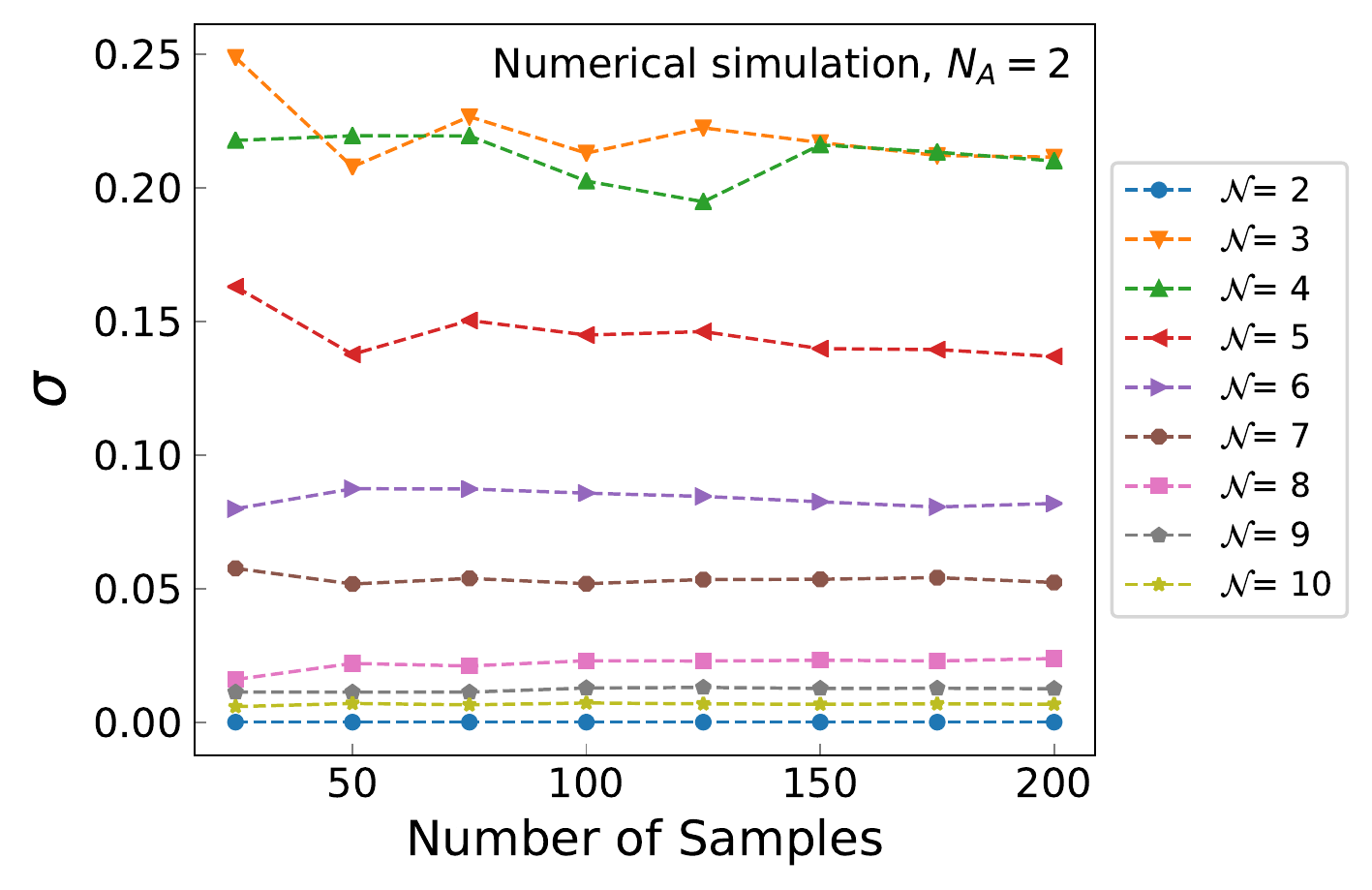} \\
  \end{tabular}

  \caption{(a) The standard deviation ($\sigma$) of ensemble average versus system size, for three different numbers of sampling,
    namely: 50, 100, and 200. (b) The standard deviation is plotted versus the number of sampling, which justifies reaching
    the reliable results for sampling larger than $\sim$100.}
  \label{errors_vs_N}
\end{figure}

The error bar that arises from the ensemble average is measured by the standard deviation ($\sigma$) of our results.
For a fixed size of A-subsystem, $N_A=2$, we have plotted the standard deviation versus system size, ${\cal N}$,
in Fig.\ref{errors_vs_N}-(a). 
The top panel also shows that the error bar decreases with the system size. This might be due to the exponential growth of Hilbert space on the system size, which results in less sensitivity to an initial configuration of light particles.
We have considered three sampling numbers, $50, 100, and 200$, for comparison in the top panel, which justifies that increasing the number of sampling does not change the final results anymore. Hence, most of our results in this work have been plotted for $200$ number of samples. It has been justified in Fig.\ref{errors_vs_N}-(b), where the standard deviation is plotted versus the number of sampling. 
The bottom panel shows almost a constant value for the error bar versus the number of sampling as far as the number of sampling is greater than $100$.

\subsection{Measurement of heavy particles}

In the next numerical experiment, we measure on the locations of heavy particles and 
obtain the reduced density matrix of light particles,
\begin{equation}
    \rho(\{ \tilde{N}_j \}) = \bra{\tilde{N}} \rho_Q \ket{\tilde{N}}.
    \label{rho_NC}
\end{equation}
Similar to the previous case, we then trace over part of light particles, namely tracing over B-subsystem, to get the reduced density matrix of light particles.
\begin{equation}
    \rho_A(\{ \tilde{N}_j \}) = \trace_B \rho(\{ \tilde{N}_j \}),
\end{equation}
The corresponding R{\'e}nyi-2  entropy of subsystem A is given as the following
\begin{equation}
 S_A^{m/M} =  - \ln [\trace_{A} \big(\rho_A(\{ \tilde{N}_j \})^2 \big) ].
 \label{S_Amm}
\end{equation}
For instance, consider an example of $N_A=2$ and ${\cal N} \geq 2$. 
After measuring the heavy particles the system will result in a pure state and the above entropy is zero, 
\begin{equation}
    S_A^{m/M} = 0  \sim L_A^{d-1},
\end{equation}
which justifies the area law for A-subsystem. 
It means that light particles are localized by the heavy ones although the heavy ones are in a thermal state.


\subsection{Quantum simulation on IBM systems }

The quantum circuit of the QDL state, which has been introduced in Sec.\ref{mqc}, is implemented on the IBM quantum simulator \cite{IBM}. However, there is a limitation due to the number of available qubits and the quantum volume of the system. As a result, we have done the simulation on the IBM system up to ${\cal N}=4$. 
The number of sampling 
on the quantum simulator is set to $20$ due to time constraints. 
Consequently, we have only considered $N_A=2$ as the size of the A-subsystem.
For measuring the Rényi-2 entropy of the circuit’s output, we estimate the final density matrix using the state tomography \cite{Smolin_2012}. As a consequence, there will be an error with respect to the classical simulation even if the state preparation, quantum gates, and measurements work ideally. In this paper, we used \textit{ibmq-oslo}, which is one of the IBM quantum resources. 
In this system, the CNOT gate has some error that its median is on the order of $O(10^-3)$.
Similarly the readout error is on order of $O(10^{-2})$.

\begin{figure}
  \centering
  \begin{tabular}{@{}l}
        \small (a) \\
    \includegraphics[width=0.9\columnwidth]{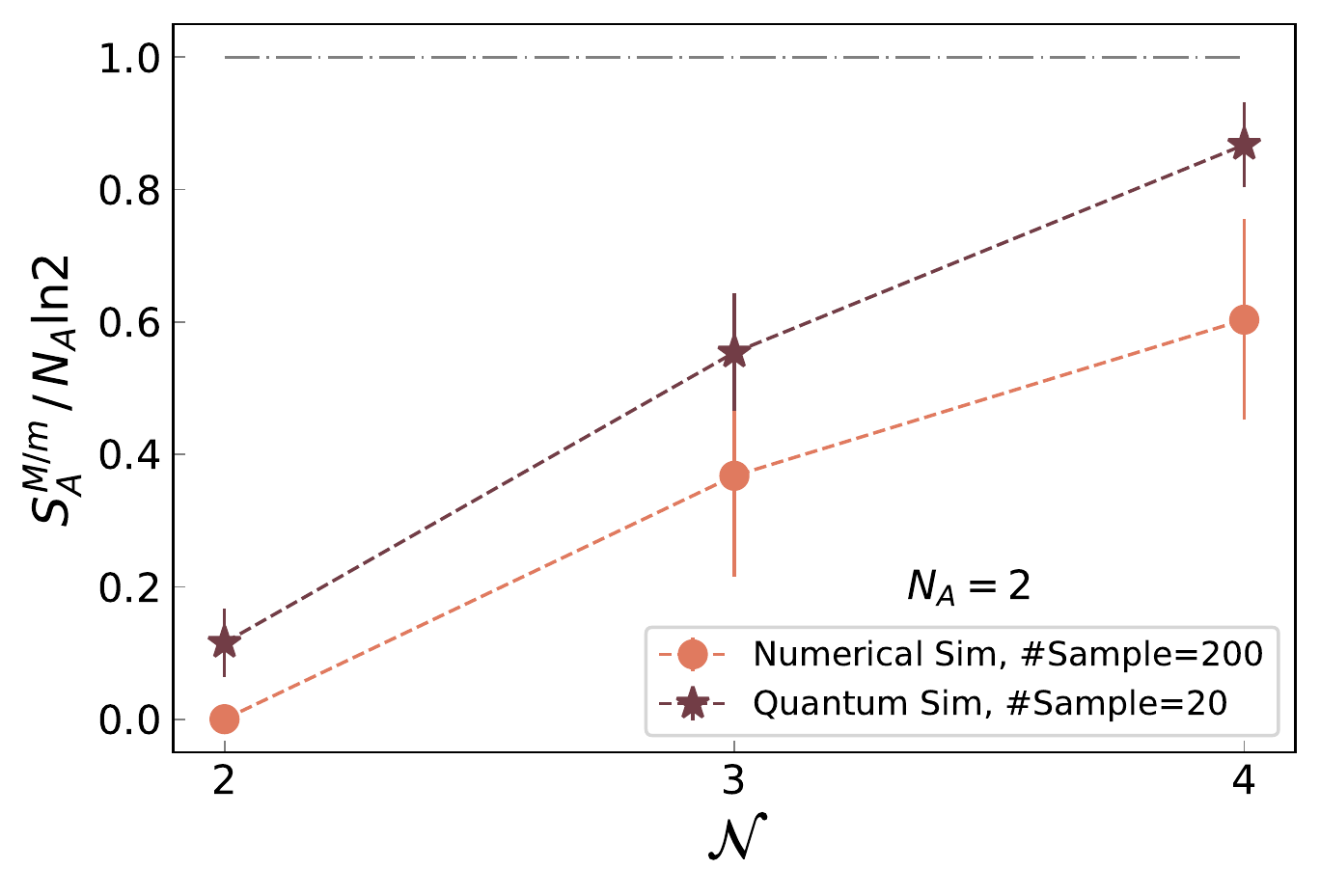} \\
  \end{tabular}\\
  \vspace{\floatsep}
  \begin{tabular}{@{}l} 
         \small (b) \\
    \includegraphics[width=0.9\columnwidth]{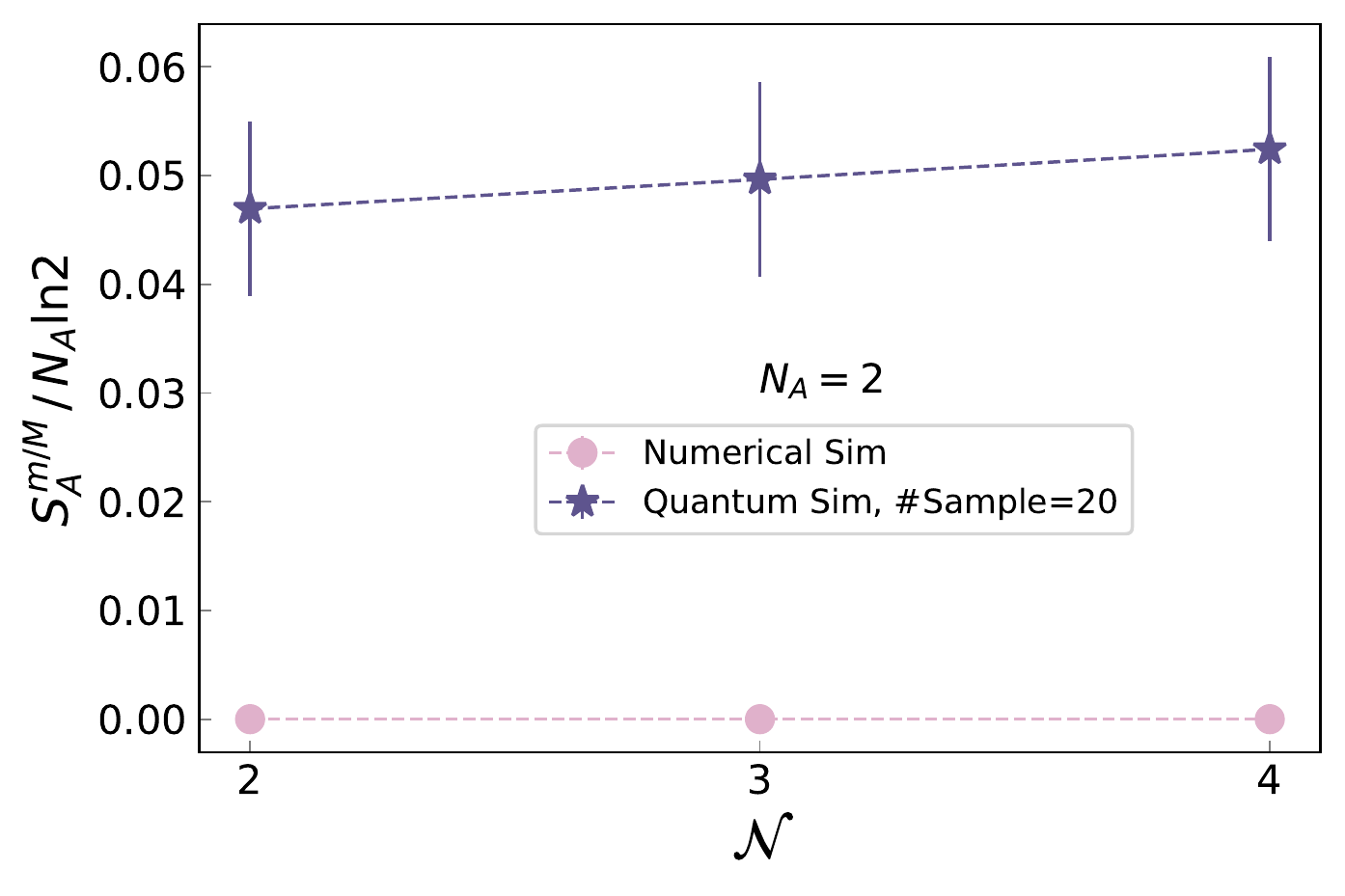} \\
  \end{tabular}

  \caption{The (a)((b)) panel shows the normalized R{\'e}nyi-2 entropy of heavy (light) particles subsystem after the measurement of light (heavy) particles is done, versus system size (${\cal N}$) for both numerical and quantum simulations. The size of the A-subsystem is fixed $N_A=2$. Quantum simulations are obtained by making measurements on IBM quantum systems, which show an offset bias due to the noise and lack of purity in the quantum simulator.
(a) Upon increasing ${\cal N}$, the entropy approaches its maximum,  $S^{M/m}_A = 2 \ln 2$, resembling the volume law for heavy particles.  (b) This figure shows the entropy of the light particles. 
For pure states, is seen in the simulations, we expect this to be zero. 
However, from the quantum simulations we get non-zero values which indicates that the states are not pure.
}
  \label{qsim_heavy_light}
\end{figure}

The results of the quantum simulations are plotted in Fig.\ref{qsim_heavy_light}. 
In Fig.\ref{qsim_heavy_light}-(a) the entropy of heavy particles $S^{M/m}$ is shown 
versus ${\cal N}=2, 3, 4$, where the location of light particles has been measured first, similar to the procedure explained in Sec.\ref{mlp}. We have also plotted the results of ideal numerical simulations for comparison. Both plots indicate the trend of volume law to reach its maximum as ${\cal N}$ increases. However, we observe the difference between numerical and quantum simulations, which is beyond the error bar of random configuration of signs (ensemble averaging).

In order to resolve the difference between numerical and quantum simulations, we obtained the 
the entropy of light particles $S^{m/M}$  versus ${\cal N}=2, 3, 4$, where the position of heavy particles has been measured first, which is plotted in Fig.\ref{qsim_heavy_light}-(b). 
As we know the quantum state for any subsystem of light particles is a pure state after measurement on heavy particles, so the entropy of light particles should be zero in an ideal simulation.
The results of numerical simulation, as expected, shows zero for the entropy of the light particles. 
However, the results of quantum simulation show a non-zero value that indicates the quantum states of the quantum simulator are not pure. 
Note that for a mixed (not pure) quantum state we get positive entropy in the measurement.
This indicates that the results of the quantum simulation are offset due to the lack of purity and presence of noise. The plot for $S^{m/M}$ in Fig.\ref{qsim_heavy_light}-(b) reflects the combined state tomography and quantum hardware error. 
For an ideal quantum simulator, we would expect to get the numerical simulation results. 

However, the trend of numerical and quantum simulations for $S^{M/m}$ is the same with a difference of less than 10 percent (within the error bar margin). It justifies that the
proposed quantum circuit (Fig.\ref{circ}) truly captures the properties of a quantum disentangled liquid state.


\section{Summary and Conclusions}
\label{conclusion}

We have simulated the quantum disentangled liquid state of Eq. (\ref{gro1}) to examine its entropy 
versus system/subsystem sizes, both numerically and on a quantum simulator. The QDL state is composed 
of light and heavy particles, where heavy ones are thermalized in a closed quantum system, while light particles 
are localized that do not thermalize. Our numerical simulation shows that the reduced density matrix of heavy particles 
obeys a volume law, which justifies their ergodic behavior. If a measurement on the light particles is done and the 
resulting entropy corresponding to a subsystem of heavy particles is calculated, it still shows volume law behavior. 
However, a measurement on the heavy particles leads to zero entropy (area law) of a subsystem of light particles.
It justifies that light particles are localized due to the measurement of heavy ones, while heavy particles are always in a
thermalized phase. We have shown that reaching the maximum entropy of heavy particles on a finite system depends
on the ratio of the system to subsystem sizes as it grows. Moreover, the error bar due to the initial configuration of light 
particles diminishes as the size of the subsystem becomes large. The average on the initial random configuration converges 
to its ensemble average after $\sim 100$ sampling. 
We have also introduced a quantum circuit to realize the QDL state on 
a quantum simulator. Our results on the quantum simulator are in agreement with our numerical simulation, which justifies
that our circuit truly demonstrates a QDL state. However, there is an offset bias between numerical and quantum simulations,
which is due to the noise in the quantum simulator. 

It will be interesting to explore an adiabatic quantum simulation of these systems \cite{biamonte2011adiabatic}. One potential candidate could be optomechanical arrays\cite{ludwig2013quantum}. 
It has been proposed to use these systems for quantum simulation of topological insulators and quench dynamics \cite{raeisi2020quench}. The optical and mechanical modes in these optomechanical arrays provide two fundamentally different degrees of freedom that can be used for simulation of the light and heavy particles.

It will also be interesting to explore further optimization of the quantum circuit. In particular, parallelization has been previously used for reducing the cost of quantum simulations \cite{raeisi2012quantum}. Similar ideas might find application for such simulations.

\section{Acknowledgements}
We acknowledge the use of IBM Quantum services for this work. The views expressed are those of the authors, and do not reflect the official policy or position of IBM or the IBM Quantum team.

\appendix
\section{Random Phase Circuit \label{rpc}} 
In this appendix we review the algorithm to break an arbitrary diagonal phase unitary gate, introduced by Schuch and Siewert \cite{Schuch}
\begin{equation}\label{psg}
      U_{\vec{\theta}} = diag \{ e^{-i\theta_0}, \cdots ,  e^{-i\theta_{2^{\cal N}-1}}\}.
\end{equation}
The building block gates of this method are one qubit z-rotation and two qubit CNOT.
\begin{equation}
      R_Z(\phi) = 
      \begin{pmatrix}
        e^{-i\phi/2 } & 0 \\
        0 & e^{i\phi/2 }
      \end{pmatrix}
      , 
      \text{CNOT} = 
      \begin{pmatrix}
        1 & 0 & 0 & 0 \\
        0 & 1 & 0 & 0 \\
        0 & 0 & 0 & 1 \\
        0 & 0 & 1 & 0 \\
      \end{pmatrix}.
\end{equation}
Suppose we prepare $\cal N$ qubits at initial computational state $\ket{\bold{x}} = \ket{x_1,\cdots, x_{\cal N}}$, $x_i \in \{0, 1 \}$. Moreover, consider a binary string of length $\cal N$, $\bold{y} = \{ y_1,\cdots, y_{\cal N} \}$. Let divide the indices that have value $1$ in the $y$ string, $\{ m| m \in \mathbb{N}, y_m = 1 \}$, with set size $M$. Then, use $M-1$ number of CNOTs with target qubit $y_{m_M}$ and $\{ y_{1}, \cdots,y_{m_{M-1}} \}$ for control qubits. Finally, if we apply a z-rotation on qubit $m_M$ with angle $\phi_{\bold{x}}$ and reuse the CNOT gates exactly the same as before, the final wave function will be as the following:
\begin{equation}
    \ket{\bold{x}} \to e^{-i\theta_\bold{x} } \ket{\bold{x}},\quad \theta_\bold{x} = \frac{1}{2} (-1)^{\bold{x}.\bold{y}} \phi,
\end{equation}
where the definition of $\bold{x}$ and $\bold{y}$ product is $\bold{x}.\bold{y} = x_1.y_1 \oplus \cdots \oplus x_{\cal N}.y_{\cal N}$. If we do the same algorithm for all $2^{\cal N}$ binary strings of $y$, the rotation angle $\theta$ is
\begin{equation}\label{pt}
    \theta_{\bold{x}} = \frac{1}{2} \sum_{y=0}^{2^{{\cal N} -1}}(-1)^{\bold{x}.\bold{y}} \phi_{\bold{x}}.
\end{equation}
More generally, since Eq. \ref{pt} is true for every computational state $\bold{x}$, we can write
\begin{equation}
    \vec{\theta} = \frac{1}{2} {\cal H} \vec{\phi},
\end{equation}
where ${\cal H}$ is the $\cal N$-bit Hadamard transformation, and ${\cal H}^2 = 2^{\cal N} \mathbb{I}$. The rotation angles $\vec{\phi} = (\phi_0, \cdots , \phi_{2^{\cal N}-1})$ for phase shift gates are related to the phases $\vec{\theta} = (\theta_0, \cdots , \theta_{2^{\cal N}-1})$ as given by
\begin{equation}\label{phi}
      \vec{\phi} = \frac{1}{2^{\cal N}-1}{\cal H}\vec{\theta}.
\end{equation}
Therefore, using CNOT and z-rotation gates, with rotation values of Eq. \ref{phi}, we are able to break a random unitary phase shift gate in Eq. \ref{psg}.

%



\end{document}